\documentstyle[preprint,prb,aps]{revtex}
\begin{document}
\draft
\title{Anisotropic two-dimensional Heisenberg model by Schwinger-boson
Gutzwiller projection method}
\author{T. Miyazaki, D.~Yoshioka, and M.~Ogata}
\address{Institute of Physics, College of Arts and Sciences, University of
Tokyo\\Komaba, Meguro-ku Tokyo 153, Japan\\}
\maketitle
\begin{abstract}
Two-dimensional Heisenberg model with anisotropic couplings
in the $x$ and $y$ directions ($J_x \neq J_y$) is considered.
The model is first solved in the Schwinger-boson mean-field approximation.
Then the solution is Gutzwiller projected to satisfy the local constraint
that there is only one boson at each site.
The energy and spin-spin correlation of the obtained wavefunction are
calculated for systems with up to $20 \times 20$ sites
by means of the variational Monte Carlo simulation.
It is shown that the antiferromagnetic long-range order remains down to
the one-dimensional limit.
\end{abstract}
\pacs{75.10.--b, 75.10.Jm, 75.50.Ee}
\section{INTRODUCTION}
Stimulated by the discovery of cuprate superconductors,
the ground state of the two-dimensional $S=1/2$
quantum antiferromagnetic Heisenberg model
has been extensively investigated by various kinds of approaches.
Although the model has never been solved exactly,
the properties of its ground state have been carefully investigated
by numerical and analytical calculations.
It has been found that ground-state energy per site is
($-0.6696 \pm 0.0004$)$J$
by series expansions \cite{series}, and ($-0.6692 \pm 0.0002$) $J$ by
Green's function Monte Carlo method\cite{GFMC}.
As for the existence of the long-range order,
it is believed that it has an antiferromagnetic long-range order.
The staggered magnetization per site is estimated to be
0.308 $\pm$ 0.008 by series expansions \cite{series} and
0.31 $\pm$ 0.01 by Green's function Monte Carlo method.\cite{GFMC} \par

On the other hand, one-dimensional $S=1/2$ quantum antiferromagnetic
Heisenberg model is solved exactly by means of the Bethe ansatz.
The exact ground-state energy per site is
$(\ln 2 - 0.25)J = -0.4431J$ and the excitation is gapless.\cite{Bethe}
Since the system is one-dimensional, it does not have a long-range order.
\par

 From these results, we expect an order-disorder phase transition
in an anisotropic Heisenberg model which
interpolates the one- and two-dimensional systems;
\begin{equation}
\label{eqn:aniH}
{\cal H} = J_x \sum_i {\bf S}_i \cdot {\bf S}_{i+\hat{x} }
       +   J_y \sum_i {\bf S}_i \cdot {\bf S}_{i+\hat{y} } ,
\end{equation}
where $i+\hat{x}$ $(i+\hat{y})$ means the site next to the $i$-th site
in the $x(y)$ direction, and $J_x=J$ and $J_y=\alpha J$ are
the coupling constants in the $x$- and $y$-direction, respectively.
Since the ground state of this Hamiltonian has no long-range order at
$\alpha=0$ (one-dimensional case)
and it has the antiferromagnetic long-range order at $\alpha=1$
(two-dimensional isotropic case),
we can expect a quantum phase transition as $\alpha$ is changed
between 0 and 1.
It had been believed that this phase boundary is at $\alpha=0$.
When $\alpha$ is infinitesimally small,
the system can be regarded as a collection of
weakly-coupled one-dimensional chains.
A simple mean-field approximation indicates that infinitesimally small
$\alpha$ makes the system ordered.\cite{2chain}
However, very recently
Parola,~Sorella and Zhong \cite{anisoPRL,anisoParola} asserted
that they found an evidence for an order-disorder transition
at finite value of $\alpha$.
They investigated the magnetization by spin wave theory and Lanczos method.
The spin wave theory predicted a breakdown of the antiferromagnetic
long-range order at about
$\alpha_c \sim 0.03(0.07)$ at first order (second order) in $1/2S$
expansions. Below this critical anisotropy,
the staggered state disappeared and disordered state was realized.
They also found that the spin wave theory generalized to finite systems
agreed with the results of the Lanczos method for $\alpha \geq 0.1$.
However, this agreement did not persist for $\alpha \leq 0.1$.
The result of the Lanczos method showed finite magnetization at $\alpha=0$.
Although they concluded the order-disorder phase transition at finite $\alpha$
from their results, it is far from conclusive.
\par

In this paper we investigate this order-disorder transition
using a different method.
Here, we consider a resonating valence bond (RVB) state \cite{RVB1,RVB2} and
adopt the Schwinger-boson description of
the spin operators.\cite{Arovas,aa,yd1,yd2}
For the isotropic two-dimensional case ($\alpha=1$),
wave functions based on the RVB singlet have been studied by many authors.
Liang, Dou\c cot and Anderson \cite{LDA} made a variational RVB wave function
with long-range bonds and found that the optimal energy per site is $-0.6688J$
and the ground state has an antiferromagnetic long-range order.
The staggered magnetization is 0.225 per site.
The variational energy is very close to the best estimated value.\cite{series}
Furthermore, recently Chen and Xiu \cite{Chen} made a trial wave function
by the Gutzwiller projection of the state obtained in the Schwinger-boson
mean-field theory. Though the wave function
has no variational parameter,
they found that it can describe the isotropic ground state
quite accurately. The energy per site is $-0.6688J$ which is
99.9 $\%$ of the best estimation and the staggered magnetization is also very
close to the best estimated value. 
In this paper, we extend the wave function of Chen and Xiu
into the anisotropic Hamiltonian for $0 \leq \alpha \leq 1$ to
investigate the order-disorder transition.
To this end, we first solve the model in the mean-field approximation.
Next, using the mean-field solution, we make variational wave functions
without doubly-occupied sites and calculate the energy and
staggered magnetization to investigate the phase transition.
\par

In section \ref{sect:mfa},
we solve the Hamiltonian in the Schwinger-boson mean-field theory.
Using this mean-field solution, we make RVB wave functions
without doubly occupied sites in section \ref{sect:makewf}.
In section \ref{sect:results}, numerical results for energies and
spin-spin correlations as a function of anisotropy
are studied by means of Monte Carlo simulations.
In the last section \ref{sect:discuss}, we discuss accuracy of our wave
function and the boundary of the order-disorder transition.

\section{Mean Field Solution}
\label{sect:mfa}
In this section we solve the Hamiltonian in the Schwinger-boson mean-field
approximation.\cite{yd1,yd2,aa}
First we introduce two kinds of bose operators $s_{i,\uparrow}$
and $s_{i,\downarrow}$ to express the spin operators,
\begin{equation}
S_i^+ = {s^\dagger}_{i,\uparrow}s_{i,\downarrow},~~{\rm and}~~
S_i^z = \frac{1}{2} (
{s^\dagger_{i,\uparrow}}
{s_{i,\uparrow}} -
{s^\dagger_{i,\downarrow}}
{s_{i,\downarrow}}).
\label{spin}
\end{equation}
The commutation relations of the spin operators $\bf{S}_i$ are satisfied
in this replacement. We impose a constraint,
\begin{equation}
\label{eqn:auxcon}
{s^\dagger_{i,\uparrow}}
{s_{i,\uparrow}} +
{s^\dagger_{i,\downarrow}}
{s_{i,\downarrow}} =1,
\label{constraint}
\end{equation}
in order to guarantee $S=1/2$.
Then the Hamiltonian is rewritten
as follows:
\begin{eqnarray}
{\cal H} &=& \sum_i \sum_z J_z({\bf S}_i \cdot {\bf S}_{i+\hat{z}}
 - {1\over4} n_in_{i+\hat{z}})
+ \mu \sum_i n_i\nonumber\\
&=& \frac{1}{2} \sum_i \sum_z \sum_\sigma J_z ({s^\dagger}_{i,\sigma}
{s^\dagger}_{i+\hat{z},-\sigma}s_{i+\hat{z},\sigma}s_{i,-\sigma}
- {s^\dagger}_{i,\sigma}
{s^\dagger}_{i+\hat{z},-\sigma} s_{i+\hat{z},-\sigma}s_{i,\sigma})\nonumber\\
&~~& + \mu \sum_i \sum_\sigma {s^\dagger}_{i,\sigma}s_{i,\sigma}~.
\label{hamilto}
\end{eqnarray}
Here
$z=x$ or $y$, $i+\hat{z}$ represents a site next to the site $i$ in the
$z$-direction,
 $n_i=
{s^\dagger_{i,\uparrow}}
{s_{i,\uparrow}} +
{s^\dagger_{i,\downarrow}}
{s_{i,\downarrow}}$, and $\mu$ is a chemical potential to be adjusted
to satisfy Eq.(\ref{constraint}) on the average.
\par
To solve this Hamiltonian in a mean-field approximation we
introduce the following mean-field parameters which give the
amplitude of the nearest-neighbor singlet pairs and an averaged occupation
number,
\begin{equation}
\Delta _z = {1\over2} \langle {s_{i,\downarrow}} s_{i+\hat{z},\uparrow}
-{s_{i,\uparrow}} s_{i+\hat{z},\downarrow} \rangle  ,
\label{rvb}
\end{equation}
\begin{equation}
n_{\sigma}=\langle s_{i,\sigma}^{\dag} s_{i,\sigma} \rangle = \frac{1}{2}.
\end{equation}
After decoupling the Hamiltonian, we rewrite the operator by its
Fourier transform:
\begin{equation}
s_{i,\sigma}= {1\over\sqrt{N}} \sum_{\bf k} {\rm e}^{i{{\bf k}\cdot
 {\bf r}}_i}
s_{{\bf k},\sigma},
\label{sk}
\end{equation}
where $N$ is the total number of lattice sites, and ${\bf k}$ are
two-dimensional wave vectors.
The mean-field Hamiltonian is then written as
\begin{eqnarray}
{\cal H}_{\rm MF} &=& \sum_{\bf k}~ [\lambda(
{s^\dagger_{{\bf k},\uparrow}}
 {s_{{\bf k},\uparrow}}
+
{s^\dagger_{-{\bf k},\downarrow}}
 {s_{-{\bf k},\downarrow}} ) +\gamma_{\bf k}
{s^\dagger_{{\bf k},\uparrow}}
{s^\dagger_{-{\bf k},\downarrow}}
+ \gamma_{\bf k}^* {s_{-{\bf k},\downarrow}}
{s_{{\bf k},\uparrow}} ]\nonumber\\
&+& 2N[ J |\Delta_x|^2 + J \alpha |\Delta_y|^2] + \frac{1}{4}N(1+\alpha) J,
\label{mfh}
\end{eqnarray}
where
\begin{eqnarray}
\lambda &=& \mu - \frac{1}{2} (1+\alpha)J ,\\
\label{lambda}
\gamma_{\bf k} &=& 2i(J \Delta_x \sin k_x + J \alpha \Delta_y \sin k_y) .
\label{gamma}
\end{eqnarray}
This Hamiltonian is diagonalized using the Bogoliubov
transformation:
\begin{equation}
{s_{{\bf k},\uparrow}} = u_{\bf k} \alpha_{\bf k} -v_{\bf k}
\beta_{\bf k}^\dagger ,
\label{sku}
\end{equation}
\begin{equation}
{s_{-{\bf k},\downarrow}} = -v_{\bf k} \alpha_{\bf k}^\dagger
+ u_{\bf k} \beta_{\bf k} ,
\label{skd}
\end{equation}
where
\begin{equation}
u_{\bf k} = {\frac{1}{\sqrt{2}}}~ [ \frac{\lambda}{{E_{\bf k}}}
+ 1]^\frac{1}{2}
\exp( \frac{i}{2}\theta_{\bf k}) ,
\label{uk}
\end{equation}
and
\begin{equation}
v_{\bf k} = {\frac{1}{\sqrt{2}}}~
[ \frac{\lambda}{{E_{\bf k}}} - 1]^\frac{1}{2}
\exp( \frac{i}{2}\theta_{\bf k}) .
\label{vk}
\end{equation}
$\theta_{\bf k}$ is the phase of $\gamma_k$,
$\gamma_k=|\gamma_k|\exp(i\theta_k)$, and
\begin{equation}
{E_{\bf k}} = \sqrt{\lambda^2 - | \gamma_{\bf k} |^2} .
\label{ek}
\end{equation}
After this transformation the Hamiltonian becomes
\begin{equation}
{\cal H}_{\rm MF} = \sum_{\bf k} {E_{\bf k}} (\alpha_{\bf k}^\dagger
\alpha_{\bf k} + \beta_{\bf k }^\dagger \beta_{\bf k})
+ {\rm const. ,}
\label{mfhd}
\end{equation}

The ground state $|G \rangle$ is defined as the vacuum of the bose operators
$\alpha_{\bf k}$ and
$\beta_{\bf k}$ : $\alpha_{\bf k}|G \rangle = \beta_{\bf k}|G \rangle=0$.
For a finite size
system, $\lambda,\Delta_x,$ and $\Delta_y$ are determined from the following
self-consistent equations,
\begin{equation}
\langle G | s_{i,\uparrow}^{\dag} s_{i,\uparrow}
    + s_{i,\downarrow}^{\dag} s_{i,\downarrow} |G \rangle =1 ,
\end{equation}
and
\begin{equation}
\langle G | s_{i,\downarrow} s_{i+ \hat{z},\uparrow}
- s_{i,\uparrow} s_{i+\hat{z},\downarrow} |G \rangle = 2 \Delta_z  ,
\end{equation}
which read
\begin{equation}
2 = {{{1\over N} \sum_{\bf k}}} \frac{\lambda}{{E_{\bf k}}}   ,
\label{eqn:an1}
\end{equation}
and
\begin{equation}
\Delta_z = {{{1\over N} \sum_{\bf k}}} \sin k_z
\frac{J_x \Delta_x \sin k_x + J_y \Delta_y \sin k_y}{E_{\bf k}} .
\label{eqn:delta1}
\end{equation}
The solution depends on the size of the system $N$, so does $u_k$ and $v_k$.
We will use the solutions for finite size systems in the following sections.

When $N$ is finite $E_{\bf k}$ never becomes zero. However, in the limit of
$N \rightarrow \infty$ it is possible that $E_{\bf k}$ vanishes at
${\bf k}={\bf K}_{\pm}= \pm(\pi/2,\pi/2)$.
In such a case it is known that we need to
introduce the Bose condensate $n_B$ \cite{yd1,yd2},
and Eqs.(\ref{eqn:an1}) and (\ref{eqn:delta1}) are rewritten as
\begin{eqnarray}
\label{an1}
2 &=& \frac{1}{(2\pi)^2} \int_{-\pi}^{\pi} dk_x  \int_{-\pi}^{\pi} dk_y
\frac{\lambda}{{E_{\bf k}}} + 2 n_B  , \\
\label{delta1}
\Delta_z &=& \frac{1}{(2\pi)^2} \int_{-\pi}^{\pi} dk_x  \int_{-\pi}^{\pi} dk_y
\sin k_z \frac{J_x \Delta_x \sin k_x + J_y \Delta_y \sin k_y}{E_{\bf k}} + n_B.
\end{eqnarray}
Here $n_B$ becomes finite only when the spectrum is gapless,
i.e. when $\lambda=2(J_x\Delta_x + J_y\Delta_y)$.
The finite value of $n_B$ means the existence of the antiferromagnetic
long-range order.
The staggered magnetization is $\sqrt{3/2}~ n_B$.

Before discussing the mean-field solution for an arbitrary anisotropy,
it is worth while to see what happens in the one-dimensional limit
where $\alpha=0$.
It is easy to see that the spectrum ${E_{\bf k}}$ cannot be gapless,
and thus $n_{\rm B}=0$.
In this case the self-consistent equations are expressed by
elliptic integrals.
By introducing $\epsilon = 2J\Delta_x/\lambda$, we can express
Eq.(\ref{an1}) and Eq.(\ref{delta1}) as
\begin{equation}
2 = \frac{1}{2\pi} \int_{-\pi}^{\pi} dk_x \frac{\lambda}{{E_{\bf k}}}
= \frac{2}{\pi}{\rm K}(\epsilon),
\label{an2}
\end{equation}
and
\begin{eqnarray}
\frac{\lambda}{J} &=& \frac{1}{2\pi} \int_{-\pi}^{\pi} dk_x \frac{\sin^2k_x}
{\sqrt{1 - \epsilon^2\sin^2k_x}}\nonumber\\
&=& \frac{2}{\epsilon^2} - \frac{2}{\pi\epsilon^2}{\rm E}(\epsilon).
\label{delta2}
\end{eqnarray}
Here K$(\epsilon)$ and E$(\epsilon)$ are the complete elliptic integrals.
By solving these equations we obtain
$\lambda = 1.379962J$, and $\Delta_x = 0.6792397$.
The excitation energy $E_k$ has a gap: $E_k \geq 0.242547J$.

For intermediate value of $\alpha$ we must solve the self-consistent equations
numerically.
The results are shown in Fig.1, where the RVB order parameters $\Delta_x$ and
$\Delta_y$, the bose condensate $n_{\rm B}$, and the energy gap $E_{\rm g}$
are plotted.
When $\alpha$ is decreased from unity, both $\Delta_y$ and $n_{\rm B}$
decreases.
The bose condensate vanishes first at $\alpha=0.1356$.
This is the point where the antiferromagnetic long-range order vanishes.
Below this value of $\alpha$, the energy gap in the spin excitation develops
quite rapidly and
$\Delta_y$ is suppressed.
$\Delta_y$ finally vanishes at $\alpha=0.1286$.
Below $\alpha= 0.1286$ the system reduces to a collection of independent
chains running parallel to the $x$-direction.
In this region the order parameter $\Delta_x$ and the gap $E_{\rm g}$ become
constant.

\section{RVB wave functions}
\label{sect:makewf}
The ground state wave function obtained in the mean-field theory
is expressed as,
\begin{equation}
\vert G \rangle=\prod_{\bf k}\exp \left
 ( -\displaystyle{\frac{v_{\bf k}}{u_{\bf k}^*}}
      s_{{\bf k},\uparrow}^{\dagger} s_{-{\bf k},\downarrow}^{\dagger} \right )
                  \vert 0 \rangle  ,\\
\end{equation}
where $\vert 0 \rangle$ is the vacuum of the Schwinger-bosons.
By the Fourier transformation for $s_{{\bf k},\uparrow}^{\dagger}$ ,
$s_{-{\bf k},\downarrow}^{\dagger}$, we get a real space representation
for this ground state;
\begin{eqnarray}
  \label{eqn:214}
\vert G \rangle&=&
 \exp [ \sum_{i,j} a_{i,j} s_{i,\uparrow}^{\dag} s_{j,\downarrow}^{\dag}]
         |0 \rangle  , \\
  \label{eqn:aij}
  a_{i,j}&=&-\frac1N\sum_{\bf k}(\displaystyle{\frac{v_{\bf k}}{u_{\bf k}^*}})
       \exp (-i{\bf k}\cdot {\bf r}_{i,j})  ,
\end{eqnarray}
where ${\bf r}_{i,j} = {\bf r}_i - {\bf r}_j$.
It is evident that the local constraint, Eq.(\ref{eqn:auxcon}),
is not satisfied in this wave function.
This is why the energy of the mean-field ground-state
is too low.\cite{foot1}
We remove this difficulty by projecting the wave function
to a space where every site is singly occupied.
Namely, we perform the Gutzwiller projection,
\begin{equation}
 \label{eqn:217}
  \vert G \rangle=P {(\sum_{i \neq j}a_{i,j}s_{i\uparrow}^{\dagger}
            s_{j\downarrow}^{\dagger})}^{\frac{N}{2}} \vert 0 \rangle .
\end{equation}
The operater {\it P} in the right hand side is the Gutzwiller projection
operater. From now on, we consider this wave function.
Since $a_{i,j}=-a_{j,i}$, the ground state of Eq.(\ref{eqn:217}) is nothing
but an RVB state which includes long-range bonds with weights $a_{i,j}$.
\par
Although it would be possible to regard every $a_{i,j}$ as a variational
parameter, we here restrict $a_{i,j}$ to be those
given in Eq.(\ref{eqn:aij}).
In the isotropic case ($J_x=J_y$), this restriction is justified by the result
itself: Chen and Xiu \cite{Chen}
have shown that this choice of $a_{i,j}$ gives excellent
results for the ground-state energy
and the staggered magnetization.
To see if Eq.(\ref{eqn:aij}) gives similarly good results for anisotropic case
or not is one of our aims in the present paper. It should be noticed that
for the anisotropic system we still have one variational parameter.
Namely, $u_k$ and $v_k$ depend on the anisotropy parameter
$\alpha=J_y/J_x$.
It is not evident that for a given value of $\alpha$
in the Hamiltonian, the same value of
$\alpha$ in the variational wave function gives the best result.
Thus we consider $\alpha$ in the mean-field equations as a variational
parameter which we denote $\alpha_{\rm p}$.


\section{Numerical results}
\label{sect:results}
In this section, we show numerical results of the variational energy,
spin-spin correlation, and staggered magnetization
for the anisotropic Heisenberg Hamiltonian.
We perform Monte Carlo simulations for lattices with various number of sites
up to 20 $\times$ 20 for the energy and the spin-spin correlation.
All the numerical calculations are performed with periodic boundary
conditions.
For each system size we solve the self-consistent equations
(\ref{eqn:an1}) and (\ref{eqn:delta1}), and calculate $a_{i,j}$ to be used
to construct the wave function at that system size.
Instead we could have used $a_{i,j}$ for an infinite size system.
However, we did not take this approach since the energy is higher
and size-scaling does not coincide with the prediction of the spin wave theory
in this case.

\subsection{Ground state}

The energy per site of the anisotropic Heisenberg model is
\begin{equation}
 \label{eqn:aveene}
   E =  J_x \langle \epsilon_x \rangle + J_y \langle \epsilon_y \rangle ,
\end{equation}
where
\begin{equation}
  \langle \epsilon_z \rangle  = \frac{1}{N} \sum_i
            \langle G | {\bf S}_i \cdot {\bf S}_{i+{\hat z}} |G \rangle  .
\end{equation}
Here $\langle \epsilon_z \rangle$, $z=x$ or $y$, is an expectation value of
the nearest-neighbor spin-spin correlation in the $x$- or $y$-direction.
These $\langle \epsilon_x \rangle$ and $\langle \epsilon_y \rangle$ depend on
the system size $N$ and the variational parameter $\alpha_{\rm p}$,
but independent of $\alpha$.
Thus we first estimate the thermodynamic limit of $\langle \epsilon_x \rangle$
and $\langle \epsilon_y \rangle$ for several values of $\alpha_{\rm p}$.
The size dependence is studied and we find the following size scaling
\begin{equation}
 \label{eqn:lsizee}
 \langle \epsilon_z (L) \rangle = \langle \epsilon_z (\infty) \rangle
 + \lambda_z L^{-3} + \cdots ,
\end{equation}
where $\lambda_z$, is a constant and $L$ is the linear dimension,
$L^2=N$.
The size-scaling coincides with the spin wave theory for a square lattice.
Then we obtain $\alpha_{\rm p}$ dependence of the energy $E$ at a fixed
$\alpha$ for the infinite size system.
Fig.~2 shows such dependence at $\alpha=0.6$ for example.
The error bars result both from Monte Carlo statistical errors and from
a fitting error of the size-scaling. It can be seen that
the optimal energy is given when $\alpha_{\rm p}=0.680$.
In a similar way, we obtain the optimal energy for various values of
$\alpha$, which is shown in Fig.~3.
When $\alpha$ is equal to zero, the system becomes the one-dimensional
$S=\frac{1}{2}$ antiferromagnetic Heisenberg chain.
In this case, the exact energy per site is given in the Bethe ansatz solution
which is $(\ln 2 - 0.25)J = -0.4431J$.\cite{Bethe}
Our variational energy per site in this case is $(-0.4337 \pm 0.0030)J$
which is 97.9$\%$ of the exact result.
This shows that our RVB wave function can also describe fairly well
the anisotropic Heisenberg Hamiltonian as well as the isotropic one.

\subsection{Staggered magnetization}
%
There are several methods to estimate the staggered magnetization.
In most numerical calculations, it is estimated from the spin-spin
correlation. Here, we use the staggered spin-spin correlation at the
longest possible distance for each direction,
\cite{review}
\begin{eqnarray}
   \label{eqn:cl2l2}
M_z^2(L)    & \equiv &  C_z (\frac{L}{2})  , \\
C_z(n) &=& [\frac{(-1)^n}{N}] \sum_i
         \langle | {\bf S}_i \cdot {\bf S}_{ i+n\hat{z} } | \rangle  ,
\end{eqnarray}
where $z=x$ or $y$ and $i+n \hat{z}$ is the $n$-th site from the site $i$ in
the $x$- or $y$-direction.
We calculate $M_z(L) $ for each lattice size and
extrapolate to the thermodynamic limit using the finite size scaling:
\begin{equation}
\label{eqn:mzl}
M_z(L)   = M_z(\infty) +  \mu L^{-1} + \cdots .
\end{equation}
where $\mu$ is a constant.
This scaling agrees with the prediction of the spin wave theory and
arguments given by Huse et al.\cite{Huse}.
The obtained $M_z(\infty)$ is the staggered magnetization.
\par
Since the variational function is a Gutzwiller projected mean-field solution,
the properties of the latter wave function is inherited by it.
The mean-field wave function which shows the long-range order
gives a finite staggered magnetization even after the Gutzwiller projection,
while that which shows no long-range order gives vanishing staggered
magnetization after the projection.
We find that, for any anisotropy, the optimized parameter $\alpha_{\rm p}$
is larger than
$0.1356$ for any $\alpha$ and, as a result, our variational state always
has a long-range order.
Furthermore we find that the magnetizations in the $x$- and $y$-directions
coincide, which is consistent with the mean-field solution.
The $\alpha$ dependence of the staggered magnetization is shown in Fig.~4.
The staggered magnetization and hence the antiferromagnetic
long-range order remains down to $\alpha=0$.

\section{DISCUSSIONS}
\label{sect:discuss}

In this paper we first solved the Hamiltonian by the Schwinger-boson
mean-field theory, and then the solution is Gutzwiller projected to obtain
the variational wave function, which we used to investigate
the order-disorder transition.
Our Gutzwiller-projected wave function is very suitable
for studying the order-disorder transition.
Since it is rewritten as a RVB state, it can represent a disordered
spin state as well as an ordered state.
As shown by Liang, Dou\c cot and Anderson \cite{LDA}, when the weight of the
singlet RVB bond, $a_{i,j}$, has a power-law behavior,
the RVB state has a long-range order.
On the other hand, when $a_{i,j}$ is short-ranged,
the RVB state becomes a disordered state with a spin gap and it has a
short range spin-spin correlation.
\par
By solving the Schwinger-boson mean-field theory for anisotropic Heisenberg
model, we construct anisotropic $a_{i,j}$ which is suitable for the
anisotropic case. Furthermore, we can compare the variational energy for the
ordered state ($\alpha_{\rm p} > 0.1356$),
and the disordered state ($\alpha_{\rm p} < 0.1356$)
by regarding $\alpha_{\rm p}$ as a variational parameter.
We have found that the optimized parameter $\alpha_{\rm p}$ is always
larger than $\alpha$ and moreover larger than $\alpha_{\rm c}=0.1356$.
(Even at $\alpha \rightarrow 0$, we get $\alpha_{\rm p}=0.205$.)
This means that the ground state is stabilized
if it develops a long-range order for all parameters, $0 \leq \alpha \leq 1$.
Although both the spin wave theory discussed by Parola et al. \cite{anisoPRL}
and the Schwinger-boson mean-field theory predict a disordered phase
for small $\alpha$, our result supports the existence of a long-range order
even for small $\alpha$.
\par

We believe the phase transition obtained in the spin wave theory and in
section  \ref{sect:mfa} at finite value of $\alpha$
is an artifact of the poorness of the mean-field solutions.
It is known that in the one-dimensional limit even though there is no
long-range order, the excitation spectrum is gapless and the spin-spin
correlation decays not exponentially but algebraically.
This algebraic decay lies in the base of our belief that an infinitesimal
coupling between one-dimensional chains causes a long-range order
in the entire system although the magnitude of the staggered magnetization
being infinitesimally small.
On the other hand our mean-field solution at $J_y=0$ gives a gapfull
excitation spectrum, and the spin-spin correlation shows an exponential decay.
This qualitative difference between the mean-field and the exact solution
does not persist in our variational wave function, since we use
$\alpha_{\rm p}> 0.205$.
We remark here that the gapfull excitation and the exponential decay of the
spin-spin correlation are the properties of the integer spin antiferromagnetic
Heisenberg chain, and the order-disorder transition in the presence of
the interchain coupling has been successfully discussed by Azzouz and
Dou\c cot \cite{Azzouz} by the Schwinger-boson mean-field theory.

Our wave function is much improved from the mean-field solution
by the Gutzwiller projection and by the optimum choice of the variational
parameter $\alpha_{\rm p}$.
However, the energy and the magnetization in the limit of $\alpha \to 0$
clearly indicates that there is  ample room for improvements of our
wave function.
Our results show discontinuity in the staggered magnetization at $\alpha=0$:
with infinitesimal $\alpha$ the magnetization jumps from zero to
$0.1227 \pm 0.023$.
This will  not be correct.
When the ordered state persists down to the limit of $\alpha = 0$,
the most natural behavior would be a continuous growth of the magnetization
from zero as $\alpha$ increases; power dependence on
$\alpha$ being quite plausible.
We expect improvements of the ground-state energy at $\alpha < 1$
will decrease the value of magnetization.
At present we do not know if such an improvement will cause the magnetization
vanish even at finite value of $\alpha$ or not.
However, our results show strong tendency of the existence of a
long-range order even for infinitesimally small $\alpha$.

In conclusion we have investigated the two-dimensional anisotropic
Heisenberg model by the Schwinger-boson Gutzwiller projection method.
Our results show existence of the long-range order in the whole range of
the anisotropic parameter $\alpha$.
This is in disagreement with the work by Parola et al.\cite{anisoPRL}
We think the order-disorder transition at finite value of $\alpha$ is
still an open question.
More elaborate work will be necessary to draw a definite conclusion for it.
We can and will try to improve the wave function by treating $a_{i,j}$ as
variational parameters.
The results will be reported in the near future.


\section*{Acknowledgements}
\vspace{0.2cm}
The authors thank H~.Yokoyama and K.~Higuchi
for helpful comments on the Monte Carlo algorithm.
This work was supported by a Grant-in-Aid for Scientific Research on Priority
Area ``Computational Physics as a new Frontier in Condensed Matter Research"
(04231105) from Ministry of Education, Science and Culture.


\newpage
\section*{FIGURE CAPTIONS}
Fig.~1. Mean-field values of order parameters $\Delta_x, \Delta_y$,
bose-condensate $n_B$ and energy gap $E_g$ as a function of $\alpha$. \\

Fig.~2.  Variational energy per site as a function of $\alpha_{\rm p}$ for
$\alpha=0.6$. The optimal energy is given at $\alpha_{\rm p}=0.680$.
\\

Fig.~3.  Energy per site as a function of $\alpha$. When $\alpha=1$,
isotropic case, the energy is $(-0.6688 \pm 0.0008)J$, and when $\alpha=0$,
one-dimensional case, it is $(-0.4337 \pm 0.0005)J$.
The exact value at $J_y=0$ is shown by an arrow.
\\

Fig.~4.  Staggered magnetization $M=M_x(\infty)=M_y(\infty)$
as a function of $\alpha$.


\begin{thebibliography}{99}
\bibitem{foot1} It is known that the energy per site is $-1.0058J$
in the isotropic case, \cite{yd1,yd2} which is about 1.5 times larger than
the best estimate value $-0.6696J$.\cite{Huse}
\end{thebibliography}
\end{document}